\documentclass[journal]{IEEEtran}

\IEEEoverridecommandlockouts
\usepackage{amsmath}
\usepackage{amssymb}
\usepackage{pifont}
\usepackage{bbding}
\usepackage{wasysym}
\usepackage{graphicx}
\usepackage{caption}
\usepackage{multirow}
\usepackage{color}
\usepackage{xcolor}

\usepackage{algorithm}
\usepackage{algpseudocode}
\usepackage{booktabs}
\usepackage{subcaption}

\newcommand{\jmyrv}[1]{\textcolor{black}{#1}}
\newcommand{\jmyrvs}[1]{\textcolor{black}{#1}}

\begin{document}
\title{Delta Debugging in the Absence of Test Oracles Through Metamorphic Testing}

\author{Mingyue Jiang, \thanks{M. Jiang is with School of Information Science and Technology, Zhejiang Sci-Tech University, Hangzhou,
China (e-mail: mjiang@zstu.edu.cn)}
Yongqiang Tian, \thanks{Y. Tian is with Department of Software Systems \& Cybersecurity, Monash University, Melbourne,
Australia (e-mail: Yongqiang.Tian@monash.edu)} %
Tsong Yueh Chen 
\thanks{T. Y. Chen is with the School of Science, Computing and Emerging
Technologies, Swinburne University of Technology, Hawthorn, VIC 3122
Australia (e-mail: tychen@swin.edu.au)}
\thanks{Corresponding author: Yongqiang Tian}
\thanks{Preprint --- under review. This is the authors' version.}
\thanks{This research was partially supported by use of the Nectar Research Cloud, a collaborative Australian research platform supported by the NCRIS-funded Australian Research Data Commons (ARDC).
It was also supported by the National Nature Science Foundation of China (Grant
No.61802349).
}
}

\maketitle

\begin{abstract}
Delta debugging provides an automatic way to minimize a program input while preserving a certain property. However, its effectiveness fundamentally relies on the availability of test oracles to determine whether a reduced input still preserves the specific property. 
Consequently, the oracle problem substantially limits the applicability of existing delta debugging techniques, particularly for oracle-deficient programs where output correctness cannot be directly determined.
To address this problem, 
this paper proposes a novel approach, DDMT, to enhance the applicability of delta debugging, especially facilitating its application to oracle-deficient programs.
Our key insight is to redesign an oracle-independent test function and incorporate it into the reduction procedure of delta debugging such that the property-preservation validation can be accomplished without requiring a test oracle.
To this end, DDMT employs the technique of metamorphic testing, which is a property-based and oracle-independent testing method.
It establishes a metamorphic testing-based test function, using it as a replacement for the original test function adopted by delta debugging.
The experiments evaluate DDMT on 66 subjects across both oracle-available and oracle-deficient scenarios, with different delta debugging approaches.
The results positively confirm that DDMT can enhance the applicability of delta debugging while often preserving or improving reduction effectiveness and query efficiency.
Furthermore, compared to the relevant delta debugging approaches, DDMT is also able to achieve performance improvements with proper configurations.

\end{abstract}


\IEEEpeerreviewmaketitle

\section{Introduction}
\IEEEPARstart{T}{est} inputs are key artifacts for program debugging.
A failure-causing input that reveals program failures can aid the debugging process in finding the bug, understanding the root cause of the bug, fixing the bug, and so on.
However, such an input may be overly large and complex, because 
it may contain some information that is irrelevant to the cause of failure.
The debugging of a program using such an input may be unnecessarily cumbersome and inefficient.
Therefore, a simplified input that preserves the failure-revealing capability of the original input can facilitate more effective debugging.

Delta debugging (DD) \cite{Zeller2002, zeller2009programs, zeller2025simplifying} is a systematic approach towards the automated input simplification.
It automatically reduces an input of the target software system into the essential minimum that still preserves the property of interest (e.g. the capability of revealing a specific type of failures).
DD is traditionally built upon the \emph{ddmin} algorithm, which follows a divide-and-conquer strategy by recursively partitioning inputs and eliminating irrelevant elements.
To simplify program inputs, the algorithm repeatedly generates reduced candidate inputs and performs property-preservation validation on each candidate. 
The algorithm continues until no further reduction is possible, and returns the smallest candidate input found that still exhibits the property of interest.

Since the inception of DD, it has attracted extensive research attention.
On the one hand, it has been extended and adapted for a wide range of applications, such as isolating cause-effect chains for failure diagnosis \cite{Zeller2002b}, simplifying method calls for failure reproduction \cite{Burger2011}, simplifying GUI event traces \cite{Clapp2016}, reducing event sequences of web applications \cite{Hammoudi2015}, isolating bug-inducing and bug-fixing changes \cite{song2024c2d2}, debugging microservice systems \cite{Zhou2018} and AI-integrated systems \cite{zhu2026}, etc. 
On the other hand, considerable research efforts have focused on improving its effectiveness and efficiency through various optimization strategies.
Recognizing that \emph{ddmin} ignores the structural characteristics of inputs, HDD \cite{Misherghi2016} exploits the hierarchical nature of program inputs and employs specific tree-based manipulators to reduce inputs. 
Similarly, GTR \cite{Herfert2017} combines \emph{ddmin} with tree transformations to support the reduction of tree-structured inputs. 
Perses \cite{Sun2018} further incorporates the formal syntax of programming languages into the reduction process, enabling more effective and syntax-aware program simplification.
Recently, ProbDD \cite{wang2021} incorporates a probabilistic model into \emph{ddmin}, in order to guide the reduction process using dynamically updated probability estimations.
WDD \cite{zhou2025wdd} adopts a different perspective by assigning weights to input elements and leveraging the weight information to optimize the reduction procedure. 
Instead of directly simplifying inputs, GReduce \cite{ren2025} reduces the execution trace of a test input generator and re-executes the generator with the reduced trace, allowing validity-preserving input reductions.

\jmyrv{
While existing techniques have made significant progress in improving reduction efficiency and minimization effectiveness, they have paid limited attention to the \emph{test oracle problem}, leaving this important problem largely unresolved.
To ensure that the reduced input preserves the property of interest, DD performs the property-preservation validation on every candidate input explored during the reduction process.
Such validation is conducted through a \emph{test function}, which requires a test oracle to differentiate between passing and failing executions, so as to identify failure-causing inputs for further reductions.
Consequently, the applicability of DD fundamentally depends on the availability of test oracles. 
Existing DD approaches are applied with explicit test oracles that exactly specify the expected behavior of the target program, including the observable failure-based oracle (i.e., program crash or a specific exit code) \cite{Zeller2002, Herfert2017, Hod2017, Groce2016}, and  output-based oracle (i.e., the expected output value) \cite{Christi2018}. 
However, reliable test oracles are often unavailable, expensive to construct, or difficult to automate in practice, causing DD to suffer from the well-known test oracle problem \cite{zeller2025simplifying, Barr2015}. 
The test oracle problem significantly limits the applicability of existing DD techniques in oracle-deficient scenarios.
}

\jmyrv{
Orthogonal to existing studies that focus on improving  effectiveness and efficiency,
this study targets the test oracle problem and aims to enhance the applicability of delta debugging without compromising its effectiveness and efficiency.
To this end, we draw inspiration from metamorphic testing (MT) \cite{Chen98, Segura16, Chen2018}, a well-established technique for alleviating the test oracle problem.
We propose DDMT, a novel delta debugging approach that systematically integrates \emph{ddmin} with MT.
The core insight of DDMT is to conduct MT rather than the conventional testing for property-preservation validation.
Specifically, DDMT proposes an MT-based test function as an alternative to the one used in conventional DD.
Moreover, it establishes systematic mappings between the original and the MT-based test functions, enabling the seamless integration of the latter into the overall reduction process. 
}

\jmyrv{
Instead of employing a test oracle, the MT-based test function leverages a metamorphic relation (MR) that encodes some necessary property of the target program.
Accordingly, the property-preservation validation is conducted via MT with respect to a specific MR. 
Consequently, DDMT no longer relies on test oracles and thus is of better applicability than DD.
On the other hand, owing to the fact that the result of property-preservation validation directly affects the partition and reduction decisions of DD, different test functions may provide varying degrees of support to the reduction process.
More specifically, due to the discrepancies in the failure-revealing effectiveness of MT and the traditional testing, DDMT may adopt different partition decisions from that of DD, exploring a distinct set of candidate inputs and ultimately producing varying results.
This further enables DDMT to exhibit better reduction performance with the support of effective MRs that have strong failure-revealing abilities.
}

We evaluate DDMT on 66 subject programs across both oracle-available and oracle-deficient scenarios.
The experiments employ two representative DD approaches: 
in the oracle-available scenario, \emph{ddmin} is employed with the use of an explicit oracle for string or textual file reduction;
in the oracle-deficient scenario, Perses \cite{Sun2018} is applied with the support of benchmark-specific oracles for program reduction.
Accordingly, DDMT and Perses$_{DDMT}$ (Perses with the integration of DDMT) are applied with the corresponding MRs in both scenarios.
The results demonstrate that although without requiring test oracles, the overall performance of DDMT is comparable to that of \emph{ddmin} in both scenarios.
Particularly, when equipped with effective MRs, DDMT exhibits promising performance, producing  12\% to 37\% smaller-sized inputs while involving 11\% to 37\% fewer test function queries for some subjects under investigation.

In summary, this study makes the following key contributions.
\begin{itemize}
    \item We propose DDMT, a novel MT-based delta debugging approach to simplifying failure-causing inputs without requiring a test oracle, enabling input reductions in both oracle-available and oracle-deficient scenarios.

    \item We extensively evaluate DDMT in different oracle situations, demonstrating that DDMT substantially improves applicability while often preserving or improving reduction effectiveness and query efficiency, and even achieves better performance with the use of effective MRs.

    \item We implement DDMT, and publicize the replication packages for conducting delta debugging on a broad range of domains.
\end{itemize}

The rest of the paper is organized as follows:
Section \ref{sec:DD} introduces the delta debugging approach.
Section \ref{sec:app} presents the details of the proposed approach.
Section \ref{sec:setup} explains the experimental setup, and Section \ref{sec:results} reports the experimental analysis results.
Section \ref{sec:dis} discusses the limitations of the proposed approach, and also clarifies the threats to validity.
Section \ref{sec:rw} presents and summarizes related studies, and Section \ref{sec:con} concludes the paper.

\section{Delta Debugging}
\label{sec:DD}

Delta Debugging (DD) \cite{Zeller2002, zeller2009programs, zeller2025simplifying} is a classic, automated approach that uses a divide-and-conquer strategy to find the root cause of a failure.
The family of DD approaches is built upon the \emph{ddmin} algorithm, which minimizes a failure-causing input via recursive input partition and validation.

\subsection{The \emph{ddmin} Algorithm}
The \emph{ddmin} algorithm aims at simplifying a program input into
a \emph{minimal} one, which is minimal in size,
 but still preserves the same failure-revealing capability as the original input \cite{Zeller2002}. 
 To this end, \emph{ddmin} conducts a series of reductions, and this reduction procedure is assisted by a test function, which performs property-preservation validation to identify failure-causing inputs, guaranteeing the preservation of the failure-revealing capability of the resulting input.

Let $t$ be an arbitrary input of the target program $P$,
and \emph{test} be the test function determining the preservation of the failure-revealing capability for inputs.
With reference to a test oracle $o$, the function \emph{test} can be described as below.
\begin{equation}
test(t, P, o) = \left\{
\begin{aligned}
F &, & if \; P(t, o) \rightarrow Fail; \\
T &, & if \; P(t, o) \rightarrow Pass.   
\end{aligned}
\right.
\end{equation}
\jmyrv{Specifically, function \emph{test} executes $P$ with $t$ by referring to $o$, in order to judge whether the execution fails or passes.
Accordingly, $t$ is reported to be failing (denoted by $F$) or passing (denoted by $T$).
A failing input reveals failures of a failed execution.
} 

The \emph{ddmin} algorithm can be described as follows.

\begin{equation}
\begin{aligned}
\forall t, test(t,P,o) = F\text{,   } & \\ddmin: (t, P, o) \rightarrow t', &\\
\text{where } & t' \in t, \text{   } test(t',P,o) = F, \\
\text{and } & \forall x \in t', test(x,P,o) \neq F
\end{aligned}
\label{eq:dd}
\end{equation}
$t'$ is a program input that is minimal in size but still preserves the same failure-revealing capability as $t$.
Starting with a given program input, \emph{ddmin} first converts the input into a set of configurable elements with respect to a certain granularity, such as characters, lines, or tokens.
After that, it processes the input via a finite number of \emph{runs}, each of which consists of the following three steps.

\begin{itemize}
\item[(1)] \emph{Input reduction via partitioning}. 
Following a partition granularity $n$ (which is initialized to $0$ and keeps updating during the input reduction procedure), the currently target input $t$ is split into $n$ inputs $t'_1,...,t'_n$ that have comparable size. %
It should be noted that due to the partition, all of the resulting candidate inputs (namely, $t'_i$, $1\le i\le n$) are of smaller size than $t$.
Next, each of the candidate inputs is validated by applying the function \emph{test}. 
If there exists at least one input, say $t'_i$, on which  \emph{test}($t'_i, P, o$) = $F$, then $t'_i$ is set as the target input in the next run by applying Step (1).
Otherwise, the complements of $t'_i$ are constructed and operated by following Step (2).

\item[(2)] \emph{Input reduction via complementing}.
For each candidate input $t'_i$, its complement input, namely, $t^{'+}_i$, is constructed with respect to $t$ ($t'_i$ $\cup$ $t^{'+}_i$ = $t$). 
If there exists at least one complement input, say $t^{'+}_i$, on which \emph{test}($t^{'+}_i, P, o$) = $F$, then $t^{'+}_i$ is further partitioned in the next run with granularity $n - 1$ according to Step (1).
Otherwise, the algorithm will change the granularity $n$ for further manipulations by following Step (3).

\item[(3)] \emph{Granularity adjustment}.
The granularity $n$ is increased to \emph{min}($|t|$, $2$$n$) (as long as $n <$ $|t|$), and then this new granularity will be applied in the next run.
\end{itemize}

\noindent
When every candidate input resulting from Steps (1) and (2) is passing, and the increasing of granularity is failed, the algorithm terminates and returns
the last failing input that has been operated.
Obviously, the smaller the resulting input, the better.
Moreover, if a smaller number of test functions is queried, then fewer candidate inputs would be handled, and thus less resources would be needed.
\jmyrv{Notably, the test function serves as a fundamental mechanism of \emph{ddmin}, whose outcome is directly utilized to make the partition and reduction decisions.
Therefore, its ability to identify failure-causing inputs will affect the reduction effectiveness and efficiency.
}

\subsection{A Motivating Example}
\begin{figure}[!t]
	\centering
		\includegraphics[width=.4\textwidth]{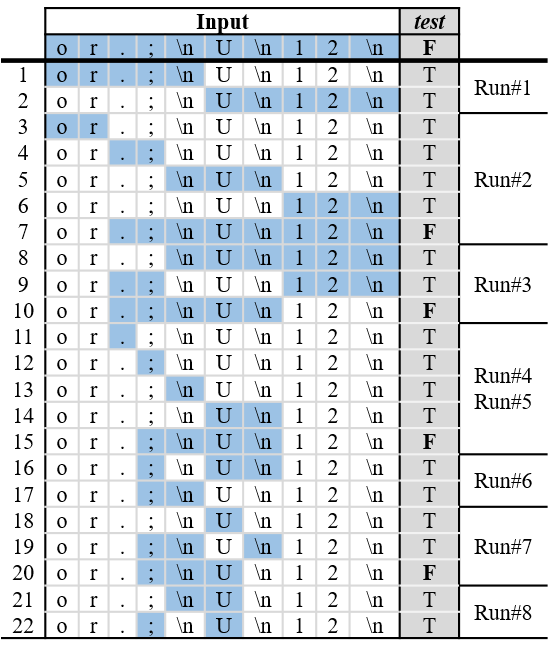}
	\caption{The detailed procedure of applying \emph{ddmin} to a faulty version of program \emph{printtokens}. The \emph{test} column reports the test outcome of individual inputs (the contents of which are highlighted in blue) in terms of passing or failing.}
	\label{fig1}
\end{figure}

For the purpose of illustration, consider the application of \emph{ddmin} to a faulty version of program \emph{printtokens} (a lexical analyzer program from the Siemens suite) as an example.
This faulty version may incorrectly handle some comments (which are texts after `;').
For a given input containing 10 characters, \emph{ddmin} finally reports a minimal failing input of size 3 after handling a sequence of 22 candidate inputs.
The detailed minimization process of \emph{ddmin} is shown in Fig. \ref{fig1}, where each row reports a candidate input and the outcome of property-preservation validation.

In run\#1, the given input is partitioned into two candidate inputs of size 5.
Since both inputs and their complements are passing (lines 1 -- 2), the granularity is increased to 4.
In run\#2, four candidate inputs are constructed and they all pass (lines 3 -- 6).
As a result, their complements are constructed and validated.
One of the complement inputs is failing (line 7).
Thus, the algorithm continues with this failing input by using a granularity of 3.
In run\#3, the three candidate inputs resulting from partitioning are already processed (as shown in lines 4 -- 6) and thus the results are directly utilized.
Their complements are further constructed,  the last one of which fails (line 10).
Run\#4 continues with granularity 2, while inputs resulting from both partitioning and complementing are already reported to be passing.
Accordingly, the granularity is increased to 4, and the relevant inputs are constructed (lines 11 -- 15), among which the first complement input fails.
In run\#6, the failing input is first partitioned into 3 candidate inputs, which are already processed and are all passing, and then the relevant complement inputs are also reported to be passing.
As a result, the granularity is increased to 4 in run\#7, where one of the complement inputs fails (line 20).
Run\#8 continues to handle this failing input with granularity 3.
In this run, all candidate inputs after partitioning and complementing are passing, and the granularity can be no longer increased for the currently target input (because $n$ is not smaller than the size of the currently target input).
As a result, the last failing candidate input (at line 20) is returned as the output.

Obviously, for \emph{ddmin}, the implementation of function \emph{test} requires a test oracle, which is a mechanism that can tell the pass or fail of individual inputs \cite{Barr2015}. 
Because of this,  the \emph{ddmin} algorithm may not be applicable when the test oracle is not available or is difficult or expensive to be applied.
In other words, the delta debugging approach faces the oracle problem \cite{zeller2025simplifying}.
More importantly, it can also be observed from the \emph{ddmin} algorithm and the illustrative example that the sequence of candidate inputs being operated is mainly determined by their testing results reported by the test function.
For example, the $7^{th}$ candidate input at line 7 of Fig. \ref{fig1} is selected for further manipulation because all candidate inputs before it are reported to be passing.
This further indicates that the failure-revealing effectiveness of the test function determines the selection of candidate inputs, and in turn affects the effectiveness and efficiency of the input reduction procedure.
Intuitively, the more failure-causing inputs can be identified earlier, the less effort will be needed to identify the final minimal failure-causing input.
Therefore, \emph{designing a new test function possessing the capability of alleviating the oracle problem as well as strong failure-revealing effectiveness can benefit the applicability, effectiveness and efficiency of DD}.

\section{Approach}
\label{sec:app}

In light of the oracle problem confronted by delta debugging and the specific strength of metamorphic testing in alleviating the oracle problem, this study proposes to systematically integrate delta debugging with metamorphic testing.
We propose DDMT, an MT-based delta debugging approach, which performs automatic input reductions without dependence on a test oracle.
This section presents the details of our approach.
We first introduce the general principles of metamorphic testing, and then explain the rationale and key mechanism of DDMT.
Finally, we revisit the illustrative example to show the specificity of DDMT.

\subsection{Metamorphic Testing}
\label{MT}

Metamorphic testing (MT)  \cite{Chen98, Segura16, Chen2018} is a testing strategy that can effectively alleviate the oracle problem.
In MT, the necessary properties of the target program or algorithm are utilized and specified as the relationships among multiple inputs and their outputs, which are known as \emph{metamorphic relations} (MRs).
Unlike traditional testing strategies that employ an oracle for checking the correctness of the output of individual inputs, MT checks whether or not multiple relevant inputs and outputs satisfy an associated MR.
Therefore, MT does not rely on a test oracle.

An MR describes the operational relationships between \emph{source inputs} and the relevant \emph{follow-up inputs}, and also encodes the expected relationships among the corresponding \emph{source} and \emph{follow-up outputs}.
A source input and its relevant follow-up inputs referring to an MR form a \emph{metamorphic group of inputs} (MG) of the MR \cite{Chen2018}.
For a \emph{violating} MG that reveals program failures, MT reports a violation of the MR (denoted by $V$).
For a \emph{non-violating} MG that fails to reveal program failures, MT reports a satisfaction of the MR (denoted by $S$).
An MR violation is an observable signal of the program failure.

\subsection{Integration of Delta Debugging with Metamorphic Testing}
\label{sec:integration}

We propose a new algorithm, DDMT, that systematically combines the \emph{ddmin} algorithm with MT.
The core insight of DDMT is to use a newly designed test function that leverages MT rather than the conventional testing technique such that failure-causing inputs can be identified without using test oracles.
Accordingly, the procedure of DDMT can be independent of a test oracle.

DDMT focuses on an input, say $t$, with which a violation of an MR $r$ can be revealed on the target program $P$.
DDMT aims to generate a minimal input $t'$ from $t$ such that the violation of $r$ is still observable. 
That is,

\begin{equation}
\begin{aligned}
\forall t, mrtest(t,P,r) = V,    & \\
 DDMT: (t, P, r) \rightarrow t', &\\
\text{where } & t' \in t, mrtest(t',P,r)=V, \\
\text{and } & \forall x \in t', mrtest(x,P,r) \neq V
\end{aligned}
\label{eq:ddmt}
\end{equation}
Notably, DDMT makes use of an MR rather than a test oracle.
Moreover, instead of applying the function \emph{test} originated from the \emph{ddmin} algorithm, it employs an MT-based test function, \emph{mrtest}, to conduct property-preservation validation.
The function \emph{mrtest} is able to identify failure-causing inputs, with \emph{mrtest}($i$, $P$, $r$) $= V$ indicating that a failure is revealed by conducting MT with an input $i$.

\subsubsection{MT-based Test Function}
\jmyrv{The core novelty of DDMT lies in a newly designed property-preservation validation function, \emph{mrtest}, which determines whether or not a violation of a specific MR can be observed on a given input. %
}
The details of function \emph{mrtest} are described as follows.

\begin{equation}
mrtest(t, P, r) = \left\{
\begin{aligned}
S &, & if \; & mt(t, P, r) \rightarrow Satisfaction; \\
V &, & if \; & mt(t, P, r) \rightarrow Violation.    
\end{aligned}
\right.
\end{equation}

For a given input $t$, function \emph{mrtest} validates it on the target program $P$ against the MR $r$.
As detailed in Algorithm \ref{alg:mrtest}, 
\emph{mrtest} conducts MT on $P$ by using $r$ and $t$, where $t$ acts as a source input for $r$.
\jmyrv{
It first constructs a follow-up input $w$ from $t$ by following $r$ (line 2).
This yields an MG, say $g = (t, w)$, for $r$.
After that, it proceeds with $g$, namely, executing $P$ on both inputs and collecting their outputs (lines 3 -- 4), and finally checking the outputs against $r$ (line 5).
If $g$ violates $r$, \emph{mrtest} returns $V$ (line 8), indicating that a violating MG can be constructed based on $t$, thereby revealing program failures by following the pipeline of MT.
Otherwise, \emph{mrtest} returns $S$ (line 6), indicating that performing MT with $t$ fails to reveal failures of $P$.
}
Obviously, function \emph{mrtest} no longer relies on a test oracle.

\begin{algorithm}
\caption{MT-based test function}
\label{alg:mrtest}
\begin{algorithmic}[1]
\Procedure{mrtest}{$t, P, r$}
   \State $w \gets r(t)$  \scriptsize // generation of follow-up input \normalsize
   \State $a \gets P(t)$  \scriptsize //source execution \normalsize
   \State $b \gets P(w)$  \scriptsize //follow-up execution \normalsize
    \If{\emph{MRCheck($a$, $b$, $r$)}}
        \State \textbf{return} \emph{S}
    \Else
        \State \textbf{return} \emph{V}
    \EndIf 
\EndProcedure
\end{algorithmic}
\end{algorithm}

\jmyrv{
According to the nature of MT,
an MR violation can be observed when either the source input or the follow-up input is failing, as elaborated in Table \ref{tab:violation}.
Therefore, the two test functions, \emph{mrtest} and \emph{test}, may have different abilities to identify inputs that can reveal program failures, and thus providing varying degrees of support to the reduction process. 
Table \ref{tab:comp} further compares possible outcomes between the two functions for the same input.
Function \emph{test} confirms the failure-revealing capability of an input $t$ only if $t$ is failing.
Differently, by treating $t$ as a source input, function \emph{mrtest} may confirm a program failure via an MR violation if 1) $t$ is failing, or 2) $t$ is not failing but the relevant follow-up input is failing.
This suggests that with the same program input, \emph{mrtest} may have a higher chance to uncover program failures, thereby guiding the reduction process to make optimized decisions.
}

\subsubsection{Correspondences and Differences between ddmin and DDMT}

\jmyrv{\emph{ddmin} and DDMT both target the minimization of a program input that reveals program failures.
They share the same reduction workflow but use different test functions to identify failure-causing inputs.}
Therefore, they have a series of correspondences, which are summarized in Table \ref{tab:corr} and further clarified as follows.

\begin{itemize}
\item \emph{An input in ddmin corresponds to an MG in {DDMT}.} 
\emph{ddmin} treats an input as a single unit for validation and manipulation, while DDMT handles each individual MG as a single unit for validation and manipulation.

\item \emph{A failing/passing input in ddmin corresponds to a violating/non-violating MG in {DDMT}}. \emph{ddmin} adopts the traditional testing, and thus an input is determined to be failing or passing. Nevertheless, DDMT applies MT, based on which an MG is reported to be violating or non-violating.

\item \emph{Validating inputs against a test oracle in {ddmin} corresponds to validating MGs against a respective MR in {DDMT}.}
For each candidate input, \emph{ddmin} checks it by referring to a test oracle, while DDMT checks the relevant MG against the respective MR.

\item \emph{Reducing an input in {ddmin} corresponds to reducing a source input of an MG in DDMT}.
\emph{ddmin} partitions an input to generate some candidate input.
Correspondingly, DDMT splits a source input into candidate source inputs, based on which candidate MGs will be constructed and manipulated.
\end{itemize}

While both approaches share the divide-and-conquer based input reduction, the use of different test functions leads to several discrepancies.
To begin with, they utilize different information.
The input to both approaches involves the target program and an input.
However, \emph{ddmin} requires the use of a test oracle, while DDMT leverages an MR.
In addition, the technical details of these two approaches are slightly different.
Function \emph{mrtest} consists of the construction of follow-up input,  the source and follow-up executions, and the checking of MR satisfaction, while function \emph{test} involves only one program execution using the given input.
Last but not least, \emph{ddmin} outputs a minimal failing input that reveals the same failures as the original input, while DDMT provides a minimal source input with which the violation of the given MR is revealed.

\begin{table}[!t]
\caption{Possible cases for an MR violation: Even if the source input is passing, an MR violation may still be observed.}
\centering
\begin{tabular}{ccc}
\toprule
\multicolumn{2}{c}{\textbf{An MG}} & \multirow{2}{*}{\textbf{Possible MT results}}\\ \cline{1-2}
Source input & Follow-up input &  \\
\midrule
\emph{Failing} & \emph{Passing} & Violation \\
\emph{Failing} & \emph{Failing} & Violation/Satisfaction \\
\emph{Passing} & \emph{Failing} & Violation \\
\bottomrule
\end{tabular}
\label{tab:violation}
\end{table}

\begin{table}[!t]
\caption{Comparison between the failure-revealing effectiveness of functions \emph{test} and \emph{mrtest}.}
\centering
\begin{tabular}{lcc}
\toprule
\textbf{An input $t$} & \textbf{\emph{test}$(t, P, O)$} & \textbf{\emph{mrtest}$(t, P, r)$} \\
\midrule
\emph{Failing} & F & V/S \\
\emph{Passing} & T & V/S \\
\bottomrule
\end{tabular}
\label{tab:comp}
\end{table}

\begin{table}[!t]
\caption{The correspondences between \emph{ddmin} and DDMT.}
\centering
\small
\setlength{\tabcolsep}{4pt}  %
\begin{tabular}{@{} p{2.0cm} p{3cm} p{3cm} @{}}  %
\toprule
\textbf{Aspects} & \textbf{\emph{ddmin}} & \textbf{DDMT} \\
\midrule
\textbf{Input} & $(t, P, o)$ & $(t, P, r)$ \\
\midrule
\textbf{Key \newline activity} &
\parbox[t]{3cm}{\raggedright $\bullet$ Validating $t$ on $P$ against the test oracle $o$ \\ $\bullet$ Reducing $t$}&
\parbox[t]{3cm}{\raggedright $\bullet$ Validating an MG with $t$ as the source input on $P$ against the MR $r$ \\
$\bullet$ Reducing $t$} \\
\midrule
\textbf{Output} &
\parbox[t]{3cm}{\raggedright A \emph{minimal} input} reveals the same failure as $t$.&
\parbox[t]{3cm}{\raggedright A \emph{minimal} input whose relevant MG violates $r$}\\
\bottomrule
\end{tabular}
\label{tab:corr}
\end{table}

\subsection{Revisiting the Motivating Example}

\begin{figure*}[]
	\centering
		\includegraphics[width=.82\textwidth]{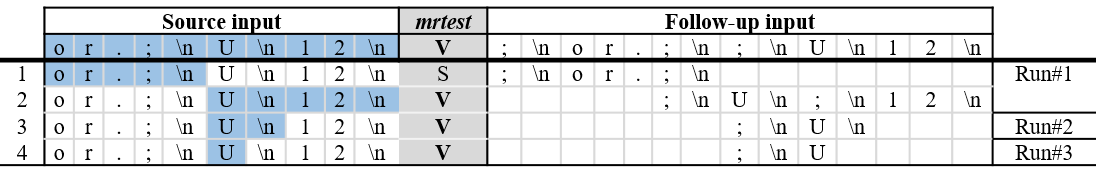}
	\caption{The procedure of DDMT for the motivating example. The \emph{mrtest} column reports the outcome of function \emph{mrtest}.}
	\label{fig2}
\end{figure*}

To explain the characteristics of DDMT,  we revisit the illustrative example as shown in Fig. \ref{fig1}.
We apply DDMT by using one MR that ``adding some comments into the text to be analyzed should not affect the analysis result because comments are ignored by printtokens".
With the same given input (whose size is 10) and the MR, DDMT outputs a result of size 1 after handling a sequence of 4 candidate inputs.
The detailed execution trace of DDMT is depicted in Fig. \ref{fig2}.
In Run\#1, the given input is partitioned into two candidate inputs, which are source inputs for constructing the relevant MGs.
It is noted that the second candidate input is passing (which is constructed and tested by \emph{ddmin} at line 2 of Fig. \ref{fig1}), while the relevant MG is violating (line 2 of Fig. \ref{fig2}).
\jmyrv{In other words, by focusing on the second candidate input, function \emph{test} identifies it as passing, while function \emph{mrtest} constructs a violating MG and thus reveals program failures.}
As a consequence, the execution trace of DDMT deviates from that of \emph{ddmin}.
That is, instead of increasing the granularity as \emph{ddmin} does, DDMT continues to partition the second candidate input in Run\#2 with granularity 2.
Again, the candidate input constructed in Run\#2, which is passing (as evidenced by the outcome of function \emph{test} as shown at line 14 of Fig. \ref{fig1}), still yields a violating MG (line 3) and thus is further partitioned in Run\#3.
As a result, the last run finds and returns a 1-sized input whose MG reveals program failures.
Obviously, DDMT is of better applicability than DD because it no longer relies on test oracles.
Moreover, due to the distinctions between the failure-revealing effectiveness of functions \emph{test} and \emph{mrtest}, \emph{ddmin} and DDMT exhibit different execution traces, yielding different debugging results.

To summarize, instead of obtaining a minimal input revealing certain failure as \emph{ddmin} does, DDMT aims at producing a minimal source input based on which the violation of a certain property is revealed.
It should be noted that with the integration, the merit of MT lies in both alleviating the oracle problem and improving the effectiveness and efficiency of DD.
As a reminder, though DDMT is originally designed to extend the applicability of \emph{ddmin} to programs without test oracles, it should be clear that DDMT can still be applied even in the presence of test oracles.

\section{Experimental setup}
\label{sec:setup}

We conducted a series of experiments by employing different delta debugging approaches and a diverse set of subject programs.
To comprehensively evaluate DDMT, our experiments consider both oracle-available and oracle-deficient scenarios.
Our experiments study the following research questions.

\noindent 
\textbf{RQ1. How well can DDMT perform as compared with \emph{ddmin} that is equipped with available test oracle?}
This RQ investigates the reduction effectiveness and efficiency of DDMT in the oracle-available scenarios.
There are two sub-questions.

\begin{itemize}
\item \emph{RQ1.1.} How does the effectiveness and efficiency of DDMT vary across the usage of different MRs?

\item \emph{RQ1.2.} To what extent can DDMT achieve performance improvements over \emph{ddmin}?
\end{itemize}

\noindent 
\textbf{RQ2. How well can DDMT perform for debugging programs facing the test oracle problem?}
This RQ investigates the reduction effectiveness and efficiency of DDMT in the oracle-deficient scenarios.

\subsection{Subjects and MRs}
\label{sec:prog}

In order to facilitate a direct comparison between \emph{ddmin} and DDMT, we selected subject programs from the Siemens suite that have available oracles. 
On the other hand, to demonstrate the capability of DDMT in the absence of test oracle,
we further selected program compilers for which no oracle is available for revealing semantic failures.

\begin{itemize}
\item \emph{Siemens programs with expected outputs as test oracles}.
Siemens suite is a widely used benchmark in software testing research \cite{Xie2013, Chen2011, Jiang2021}. 
It consists of several small to medium-sized C programs, and each subject is accompanied by a correct version and multiple faulty versions with seeded defects. 
For these programs, the correct version can provide the expected output as a test oracle.
We selected subject programs where the input contains either \emph{String} or \emph{file} parameters so that the token-based input reduction is feasible.
We excluded subject programs for which the input is unable to be reduced.

\item \emph{Program compilers without available test oracles.}
We further selected subjects from previous studies on delta debugging \cite{Sun2018, wang2021, zhou2025wdd, zhang2025toward}, where each subject consists of a program compiler and an input C program.
There are mainly two categories of subjects, those demonstrating an observable failure of the compiler (e.g., the compilation of the input program ends with a crash or abnormal exit code), and those ending with a silent failure without any observable signals (i.e., the input program is compiled successfully but the resulting executable deviates from the expected program semantics).
In the latter case, a test oracle is unavailable, and thus DD is unable to be applied without some auxiliary support.
\end{itemize}

To apply DDMT, we reused MRs from existing publications in order to avoid any biases.
For each subject program from the Siemens suite, we adopted three MRs (which are respectively referred to as MR1, MR2, and MR3 throughout the rest of this paper).
\jmyrvs{The MRs for \emph{printtokens} modify the input text by changing character cases and transforming certain tokens into identifiers or numeric tokens \cite{Xie2013, Jiang2021}. The MRs for \emph{printtokens2} transform specific tokens into identifiers, and also manipulate comments \cite{Xie2013, Jiang2021}. The MRs for \emph{replace} focus on regular expressions, including brackets, escape characters, and specific regex patterns \cite{Chen2011, Xie2013}. The MRs for \emph{schedule} involve substituting the quantum-expire command, the adding job command, and the block, unblock commands \cite{Chen2011}.}
For C compilers, we adopted one existing MR, EMI \cite{Le2014}, which transforms the original C program to generate equivalent variants, and expects all of them to produce the same outputs for a given set of test inputs.  

To facilitate comparisons between \emph{ddmin} and DDMT, we kept only subject programs for which both approaches successfully reveal failures with a certain set of same inputs.
As a result, a total of 66 faulty programs were selected, including 58 faulty programs from \emph{replace}, \emph{printtokens}, \emph{printtokens2}, 
and \emph{schedule} of the Siemens suite, and 8 clang/gcc compilers exhibiting silent failures.
The information of these subjects are summarized in Table \ref{tab:subjects}.

\subsection{Evaluation Metrics}
Both \emph{ddmin} and DDMT target producing a minimal failure-causing input (the former results in a failing input, while the latter ends with an input that is the source input of a violating MG).
According to this, their effectiveness mainly refers to  the \emph{size} of the resulting input.
On the other hand, from the perspective of efficiency, it is reported that the time spent on querying the test function dominates about 98\% of the total execution time of reduction procedure \cite{Herfert2017}.
Then, a smaller \emph{number of queries} (namely, the number of \emph{test} or \emph{mrtest} function invocations) indicates a better efficiency.
Therefore, in this study, we followed existing studies \cite{Zeller2002, wang2021, Misherghi2016, Sun2018} to use three metrics, the \emph{size}, the number of \emph{queries}, and the \emph{time cost}.

\begin{table}[!t]
\caption{Subject programs and test inputs.}
\centering
\begin{tabular}{lccccc}
\toprule
\textbf{Subject} & \textbf{\#Programs} & \textbf{\#Failing inputs} & \multicolumn{1}{c}{\textbf{Oracle Situation}} \\
\midrule
printtokens       & 7  & 525  & \multirow{4}{*}{\parbox[t]{2cm}{With the expec-\\ted output as an \\ oracle}} \\
printtokens2      & 10 & 2430  &                       \\
replace           & 32 & 2100  &                       \\
schedule          & 9  &  766 &                       \\
\midrule
clang/gcc         & 8  & 8 & Oracle unavailable                    \\
\bottomrule
\end{tabular}
\label{tab:subjects}
\end{table}

We measured the size of inputs of Siemens programs using the number of characters, and measured the size of inputs of compilers using the number of tokens.
Furthermore, we collected the number of invocations of function \emph{test} for \emph{ddmin}, and collected the number of invocations of function \emph{mrtest} for DDMT.

\subsection{Implementations}

We directly applied \emph{ddmin} and DDMT to Siemens programs, for conducting the token-level input reduction.
For program compilers, we selected the state-of-the-art delta debugging technique for program reduction, Perses \cite{Sun2018}.
We further built Perses$_{DDMT}$ by adapting DDMT to Perses, using MR violations instead of manually specified oracles to decide whether a candidate program should be preserved during reduction.

\begin{itemize}
	\item \textbf{\emph{ddmin} and DDMT.} 
	We adopted the implementation of \emph{ddmin} proposed by the original authors (https://www.st.cs.uni-saarland.de/dd/) \cite{Zeller2002}. 
	With reference to the oracle situations of Siemens programs, we implemented a function \emph{test} to support \emph{ddmin}.
	The function \emph{test} runs a given input on the target program, and determines the pass or fail by comparing the actual output with the expected output, which is obtained via running the correct version of the target program with the same input.
	We implemented DDMT upon the implementation of \emph{ddmin}.
	Specifically, we implemented each of our MRs, for supporting the automated follow-up inputs generation as well as the MR checking of source and follow-up outputs.
	Furthermore, we implemented function \emph{mrtest}, which conducts MT on the target program using a given source input and the MR, and provides a satisfaction or violation as an output.

\item \textbf{Perses and Perses$_{DDMT}$}. 
        We adopted the implementation of Perses \cite{Sun2018}.
        Specifically, to apply Perses to subjects having the test oracle problem, we re-implemented the function \emph{test} by employing benchmark-specific oracles.
        Moreover, we implemented Perses$_{DDMT}$ based on a \emph{mrtest} function and the implementation of EMI MR.

\end{itemize}

Our experimental equipment is an operating system with 8-core 16-thread Intel(R)Core(TM)i7-10700 CPU (2.90GHZ), 16 GB RAM, and Ubuntu Linux 16.04.

\section{Results and Analysis}
\label{sec:results}

\subsection{Evaluation results of Siemens programs}

In our experiments on Siemens programs, \emph{ddmin} successfully reduces 5,821 failing inputs, and DDMT successfully reduces 44,097 inputs involved in violating MGs.
When applying DDMT to Siemens programs, three MRs were utilized for each subject program.
Therefore, we first report the performance of DDMT with respect to different MRs, based on which MRs exhibiting the best and worst performance for each subject program can be identified.
Then, we conduct comparison analysis between \emph{ddmin} and DDMT with the usage of the best and worst MRs, respectively.

\begin{figure}[!t]
    \centering
    \begin{subfigure}[t]{\linewidth}
        \centering
        \includegraphics[width=\textwidth]{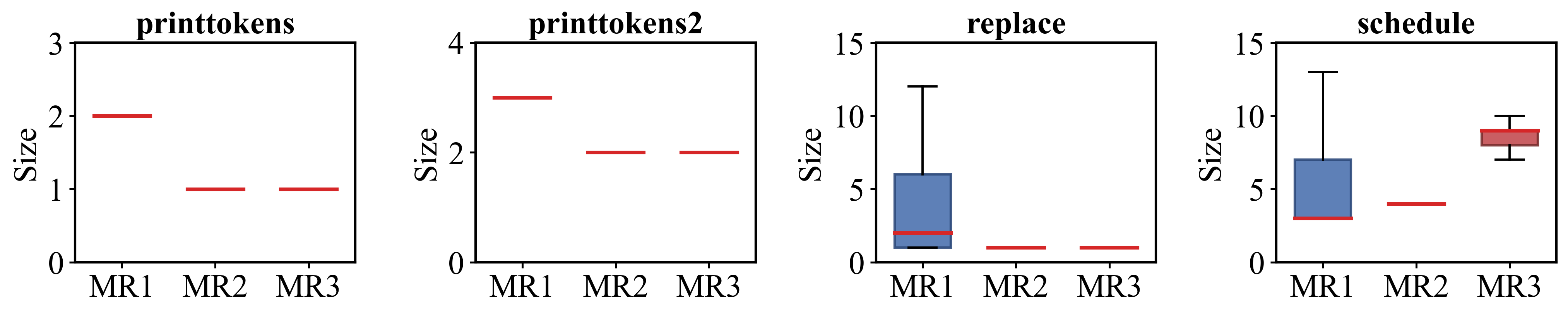}
        \caption{Size.}
    \end{subfigure}
    \hfill 
    \begin{subfigure}[t]{\linewidth}
        \centering
        \includegraphics[width=\textwidth]{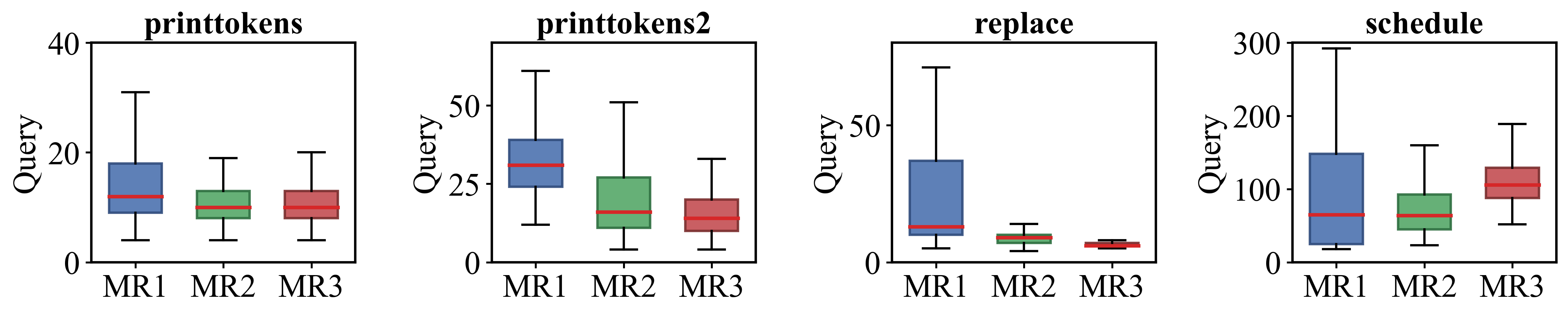}
        \caption{Query.}
    \end{subfigure}

    \hfill 
    \begin{subfigure}[t]{\linewidth}
        \centering
        \includegraphics[width=\textwidth]{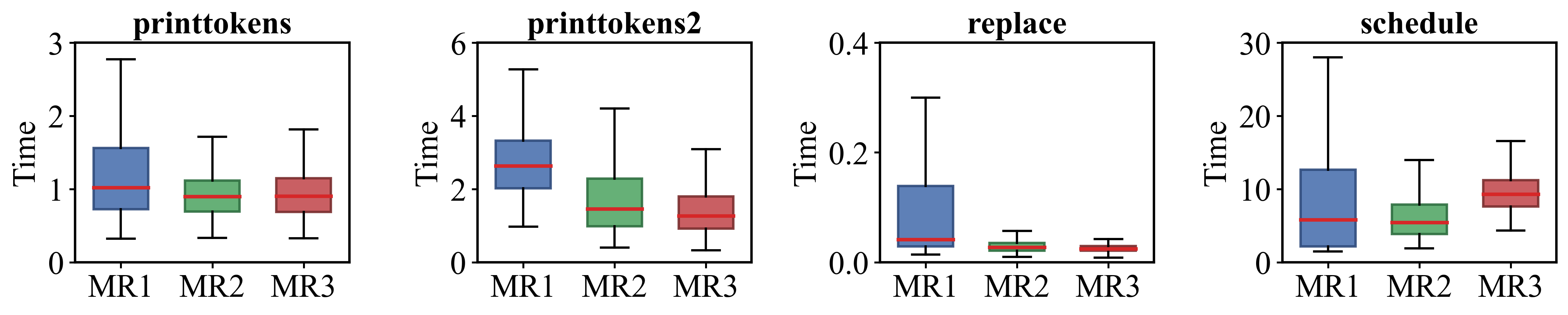}
        \caption{Time.}
    \end{subfigure}
    \caption{Performance of DDMT with three different MRs on Siemens programs.}
    \label{fig:rq1-1}
\end{figure}

\begin{table}[!t]
\caption{Comparison of the performance of DDMT with reference to different MRs.
For two MRs $A$ and $B$, $A \approx B$ indicates that two MRs are of comparable performance (\emph{p}-value $\ge$ 0.05), while $A \gg B$ means that $A$ is of better performance than $B$ (\emph{p}-value $<$ 0.05). }%
\centering
\setlength{\tabcolsep}{4pt}  %
\begin{tabular}{@{} p{1.5cm} p{2.2cm} p{2.2cm} p{2.2cm} @{}}%
\toprule
\textbf{Subjects} & \textbf{Size} & \textbf{Query} & \textbf{Time} \\
\midrule
printtokens   & \scriptsize  MR3$\approx$MR2$\gg$MR1 & \scriptsize MR3$\approx$MR2$\gg$MR1 & \scriptsize MR3$\approx$ MR2$\gg$MR1 \\
\addlinespace
              & \multicolumn{3}{l}{\footnotesize Best: MR2 \; Worst: MR1} \\
\midrule
printtokens2  & \scriptsize MR3$\gg$ MR2$\gg$MR1 & \scriptsize MR3$\gg$MR2$\gg$MR1 & \scriptsize MR3$\gg$MR2$\gg$MR1 \\
\addlinespace
              & \multicolumn{3}{l}{\footnotesize Best: MR3 \; Worst: MR1} \\
\midrule
replace       & \scriptsize MR3$\gg$ MR2$\gg$MR1 & \scriptsize MR3$\gg$MR2$\gg$MR1 & \scriptsize MR3$\gg$MR2$\gg$MR1 \\
\addlinespace
              & \multicolumn{3}{l}{\footnotesize Best: MR3 \; Worst: MR1} \\
\midrule
schedule      & \scriptsize MR1$\gg$MR2$\gg$MR3 & \scriptsize MR1$\approx$MR2$\gg$MR3 & \scriptsize MR1$\approx$MR2$\gg$MR3 \\
\addlinespace
              & \multicolumn{3}{l}{\footnotesize Best: MR1 \; Worst: MR3} \\
\bottomrule
\end{tabular}
\label{tab:mr-comp}
\end{table}%

\subsubsection{Performance of DDMT with respect to different MRs}
Fig. \ref{fig:rq1-1} reports the performance of DDMT in terms of
the reduction effectiveness (the size of the resulting inputs) and efficiency (the number of queries and the time cost), respectively.
A boxplot contains three boxes, each of which shows the distributions of the data for an MR.
In a box, the red line inside it represents the median value of the data, and the ends of the box denote the upper and lower quartiles.
It can be visually observed that for each of the subject programs, both effectiveness and efficiency of DDMT vary substantially with the use of different MRs.

\begin{table*}[!t]
\centering
\caption{Overall comparison results of \emph{ddmin} and DDMT on Siemens programs. The \emph{average} size, number of queries and time are reported. $\uparrow$ represents the improvements achieved by DDMT, which is calculated by (X - Y)/X (X and Y respectively denote the values from \emph{ddmin} and DDMT with respect to an evaluation metric).}

\setlength{\cmidrulekern}{0.5cm}
\subfloat[Comparison between \emph{ddmin} and DDMT using the best MR]{%
\centering
\begin{tabular}{c| c c c c @{\hspace{0.5cm}} c c c c @{\hspace{0.5cm}} c c c c}%
\hline
 \multirow{2}{*}{\textbf{Subjects}}   & \multicolumn{4}{c}{\textbf{Size}} & \multicolumn{4}{c}{\textbf{Query}}  & \multicolumn{4}{c}{\textbf{Time (seconds)}} \\ \cmidrule(r){2-5} \cmidrule(r){6-9} \cmidrule(r){10-13}    
 
 & \emph{ddmin} & DDMT & \emph{p-value} & $\uparrow$ & \emph{ddmin} & DDMT & \emph{p-value}& $\uparrow$ & \emph{ddmin} & DDMT & \emph{p-value}& $\uparrow$ \\  \hline                
 
printtokens  &3.72&2.43&4.32e-08 & 35\%  & 30& 19 &5.00e-09 &37\%  &0.08 &1.68&2.81e-19 & $>$-200\% \\ 
printtokens2 &1.96&2.24&0.0003&-14\%     &21&22&0.1103 &-5\%    &0.06 &1.97 &2.55e-29&$>$-200\% \\

replace   &3.28&2.89&0.0013&12\%    &19&17&0.0006&11\%     &0.13&0.19&0.8190&-46\% \\

schedule  &14.70&9.30&0.0148&37\%    &221&162&0.1055&27\%   &1.36&13.94&0.002&$>$-200\% \\ \hline
\end{tabular}
}
\vspace{4pt}

\subfloat[Comparison between \emph{ddmin} and DDMT using the worst MR]{%
\centering
\begin{tabular}{c| c c c c @{\hspace{0.5cm}} c c c c  @{\hspace{0.5cm}} c c c c}%
\hline
 \multirow{2}{*}{\textbf{Subjects}}   & \multicolumn{4}{c}{\textbf{Size}} & \multicolumn{4}{c}{\textbf{Query}}  & \multicolumn{4}{c}{\textbf{Time (seconds)}} \\ \cmidrule(r){2-5} \cmidrule(r){6-9} \cmidrule(r){10-13}     
 
 & \emph{ddmin} & DDMT & \emph{p-value} & $\uparrow$ & \emph{ddmin} & DDMT & \emph{p-value}& $\uparrow$ & \emph{ddmin} & DDMT & \emph{p-value}& $\uparrow$ \\  \hline              
 
printtokens   &2.29&2.54&9.54e-06 &-11\%   &27&23&0.0003&15\%  &
0.09&2.05&1.44e-26&$>$-200\% \\ 

printtokens2    &2.00&2.94&7.99e-75&-47\%   &24&34&5.76e-53& -42\%    &0.06&2.98&3.73e-82&$>$-200\% \\

replace   &2.91&4.36&2.46e-10&-50\%    &19 &29&8.91e-11&-53\%     &0.05&0.15&3.68e-31&$>$-200\% \\
schedule   &14.05&8.47&6.79e-28&40\%    &204&116&2.32e-24&43\%     &1.31&10.16&2.94e-33& $>$-200\%\\ \hline

\end{tabular}
}
\label{tab:best}
\end{table*}

For each subject program, we further conducted the Wilcoxon rank-sum test \cite{riina2023} on the data from individual pairs of MRs, to check whether or not there is a significant difference between them.
The results are summarized in Table \ref{tab:mr-comp}.
Among 12 comparisons on the \emph{size} of the resulting inputs, 11 pairs of MRs exhibit significant differences (\emph{p}-value $<$ 0.05).
Meanwhile, 10 pairs of MRs show significant differences in terms of the number of queries, and significant differences in time cost are also observed in 10 pairs of MRs.
These results indicate that both the effectiveness and efficiency of DDMT are highly sensitive to the used MRs.
This further consolidates the observation that MR is a key factor to the performance of MT-based approaches, as previously reported by prior studies \cite{Liu2013, li2025}.

\jmyrv{Based on the comparison results, we ranked the three MRs, and then identified the MRs with the best and worst performance with respect to each subject program.
The results are reported in Table \ref{tab:mr-comp}.
It can be observed that the ranks of MRs are quite similar across different metrics.
Moreover, the performance of the best MR (ranked first) and the worst MR (ranked last) is always statistically different.
This results suggest that a good MR can consistently achieve better reduction effectiveness with less resource consumptions.}

\subsubsection{Comparison between ddmin and DDMT with the use of the best and worst MRs}

We then compared the performance of \emph{ddmin} with that of DDMT.
As aforementioned, DDMT shows varying performance when different MRs are utilized.
Therefore, our comparison respectively analyzed DDMT with the best and the worst MRs (as reported in Table \ref{tab:mr-comp}).

As a reminder, \emph{ddmin} focuses on reducing a failing input, while DDMT focuses on reducing a source input whose MG of the given MR is violating.
What's more, as elaborated in Section \ref{sec:integration}, a violating MG may not necessarily have a failing source input.
As a result, the application of these two approaches may involve different target programs and inputs, and the application of DDMT using different MRs may also involve different target programs and inputs.
Therefore, in order to fairly compare \emph{ddmin} and DDMT (with an MR), we collected and analyzed the experimental results for a set of programs and inputs to which both approaches are successfully applied.
That is, each comparison focuses on a target program and an input, which is associated with two groups of data (including the size, the number of queries, and the time cost) from \emph{ddmin} and DDMT, respectively.

\begin{figure}[!t]
    \centering
    \begin{subfigure}[t]{\linewidth}
        \centering
        \includegraphics[width=0.9\textwidth]{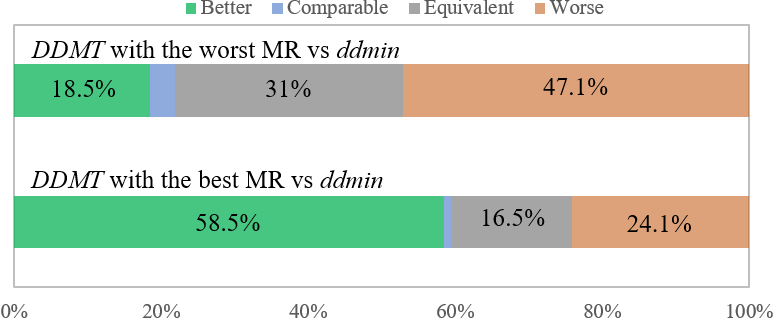}
        \caption{Comparison of the input size and the number of queries.}
    \end{subfigure}
    \hfill 
    \begin{subfigure}[t]{\linewidth}
        \centering
        \includegraphics[width=0.9\textwidth]{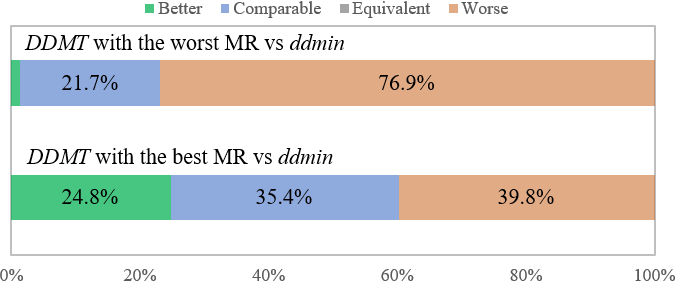}
        \caption{Comparison of the input size, the number of queries, and the time cost.}
    \end{subfigure}
    \caption{Detailed comparison results of DDMT and \emph{ddmin} on the Siemens programs.}
    \label{best-worst}
\end{figure}

A total of 1,577 comparisons were conducted, including 528 comparisons on \emph{ddmin} and DDMT using the best MR, and 1,049 comparisons on \emph{ddmin} and DDMT using the worst MR.
We performed the Wilcoxon signed-rank test \cite{riina2023} to statistically compare individual groups of data.
Tables \ref{tab:best} summarizes the overall performance of the two approaches by reporting the mean value, the statistical test result, and the improvements achieved by DDMT, in terms of each of the evaluation metrics.
Additionally, Fig. \ref{best-worst} presents the details of comparison, namely, the percentage of comparisons where DDMT is \emph{better} than, \emph{worse} than, \emph{comparable} to, or \emph{equivalent} to \emph{ddmin}.
In a comparison, DDMT is regarded to perform \emph{better} (\emph{worse}) than \emph{ddmin} if it yields better (worse) results with respect to all of given metrics;
it is \emph{comparable} to \emph{ddmin} if it yields a better result for some metrics while yielding worse results for the others;
and it is \emph{equivalent} to \emph{ddmin} if they produce the same results for all of the given metrics.

The comparison between \emph{ddmin} and DDMT with the usage of the best MR shows that DDMT often 
delivers positive improvements.
As reported in Table \ref{tab:best} (a), DDMT yields reduced inputs with smaller size than \emph{ddmin} and achieves a statistically significant improvement in simplification effectiveness (\emph{p-value} $<$ 0.05), ranging from 12\% to 37\% across three out of the four subject programs.
With regard to the efficiency, DDMT involves a comparable number of test function invocations to \emph{ddmin} on \emph{printtokens2}, while achieving improvements of 11\% to 37\% on the other three subject programs.
Nevertheless, it always incurs higher time cost as compared with \emph{ddmin}.
Apart from the subject \emph{replace} where DDMT exhibits a very tiny differences in time cost (0.19 vs 0.13 seconds), it requires much more time than \emph{ddmin} to accomplish the task of input reduction.
\jmyrv{
However, according to the detailed analysis result as presented in Fig. \ref{best-worst}, DDMT is better than \emph{ddmin} in 58.5\% of cases with reference to the input size and the number of queries (Fig. \ref{best-worst} (a)). %
When considering all of the three metrics, it performs better in 24.8\% of cases, and it is comparable to \emph{ddmin} in 35.4\% of cases (Fig. \ref{best-worst} (b)).
}
To summarize, with the support of good MRs, DDMT offers better simplification effectiveness and efficiency for the majority of programs and inputs, but overall incurring higher time overhead.

In Table \ref{tab:best} (b), the comparison results between \emph{ddmin} and DDMT with the usage of the worst MR are reported.
DDMT generally performs well on the subject program \emph{schedule}, with positive improvements in both the size and the number of queries.
On another subject program, \emph{printtokens}, it yields inputs of larger size, while requiring a smaller number of test invocations.
It is also observed that DDMT exhibits an explicit performance decrease on the other two subject programs, and it also incurs higher time overhead than \emph{ddmin} on all of the subject programs.
When inspecting the details of comparison results reported in Fig. \ref{best-worst}, \jmyrv{it is observed that DDMT performs no worse than \emph{ddmin} in about 60\% of cases in terms of both the input size and the number of queries (Fig. \ref{best-worst} (a)), while it yields better or comparable results in about 25\% of cases when considering all three metrics.}
These results suggest that when ordinary or even ineffective MRs are applied, DDMT can still retain comparable performances to \emph{ddmin} for a quite large number of programs and inputs.

In summary, the application of \emph{ddmin} and DDMT to the Siemens programs shows that DDMT is able to achieve promising effectiveness and efficiency as compared with \emph{ddmin}.
On one hand, the discrepancies between the results from DDMT and \emph{ddmin} confirm that an oracle-based test function and an MR-based test function will guide the reduction procedure in different ways, leading to varying reduction results.
On the other hand, although the performance of DDMT varies across different MRs, it exhibits significant performance improvements with the use of good MRs, and it still delivers better or comparable results for a considerable number of subjects even with an ordinary or ineffective MR. 
The results show that DDMT improves applicability while often preserving or improving reduction effectiveness and query efficiency.
These results further suggest that in the scenario where a test oracle is available, although \emph{ddmin} is applicable, it is still worthwhile to use DDMT with properly identified MRs so as to enhance the minimization effectiveness as well as reduction efficiency.

\subsection{Evaluation results of program compilers}

We apply Perses and Perses$_{DDMT}$ to reduce input programs for gcc and clang compilers, where silent failures occur without observable signals.
In this oracle-deficient scenario, 
Perses was configured with explicit benchmark-specific oracles, whereas Perses$_{DDMT}$ was configured with an MR.

The detailed results are reported in Table \ref{tab:compiler}. 
Averagely, both approaches exhibit comparable performance across all metrics.
Perses$_{DDMT}$ demonstrates a slight improvement in input size (368 vs 369 tokens, a 0.1\% decrease) and the number of queries (4623 vs 4666 queries, a 0.9\% decrease), while incurring slightly higher time overhead (7,385 vs 7,351 seconds, a 0.5\% increase).  

\begin{table*}[!t]
\caption{Comparison results of Perses and Perses$_{DDMT}$ on program compilers. }%
\centering
\begin{tabular}{c| c c c @{\hspace{0.5cm}} c c c  @{\hspace{0.5cm}} c c c}%
\hline
 \multirow{2}{*}{\textbf{Subjects}}   & \multicolumn{3}{c}{\textbf{Size}} & \multicolumn{3}{c}{\textbf{Query}}  & \multicolumn{3}{c}{\textbf{Time (seconds)}} \\ \cmidrule(r){2-4} \cmidrule(r){5-7} \cmidrule(r){8-10}     
 
 & Perses & Perses$_{DDMT}$ & $\uparrow$ &Perses & Perses$_{DDMT}$ & $\uparrow$ &Perses & Perses$_{DDMT}$ & $\uparrow$ \\  \hline                
 
clang-21582    &      559   &    559    & 0.0\%  &     6468   &   6457   &  0.2\%  &     10140  &   10117    & 0.2\% \\ 

clang-23309    &      475    &   475     &0.0\%   &   2167    &  1959     & 9.6\%    &  8404     & 8334     &0.8\% \\
clang-26350    &      196    &   194     &1.1\%    &  5374    &  5539     &-3.1\%     &  7709     & 8204     &-6.4\% \\
clang-27137    &      268     &  268     &0.0\%    & 10787    & 10791    & 0.0\%     &  9764    &  9752     &0.1\% \\
clang-31259    &      384     &  384     &0.0\%    &  2480    &  2481     &0.0\%     & 5055     & 5006     &1.0\% \\
gcc-66186      &     327      & 327     &0.0\%     & 2845     & 2666     & 6.3\%      & 5355     & 5347     &0.2\% \\
gcc-66375      &       440     &  440     &0.0\%    &  3469   &   3355    & 3.3\%      & 6784     & 6745     &0.6\% \\
gcc-70127      &       301     &  301     &0.0\%     & 3736    &  3734     & 0.1\%      & 5600     & 5578     &0.4\% \\ \hline

Average  &     369      & 368     &0.1\%   &  4666    &  4623    & 0.9\%  &    7351   &   7385   &  -0.5\% \\ \hline
\end{tabular}
\label{tab:compiler}
\end{table*}%

For minimization effectiveness, Perses and Perses$_{DDMT}$ consistently yield programs of identical size in 7 out of 8 cases, and Perses$_{DDMT}$ yields a smaller-sized program in the remaining one case.
When inspecting the number of queries, Perses$_{DDMT}$ generally needs fewer test function invocations than Perses, as evidence by the 0.9\% improvements achieved across all cases.
Particularly, Perses$_{DDMT}$ shows explicit improvements in three cases, by requiring 9.6\%, 6.3\%, and 3.3\% fewer queries.
These results confirm the input reduction capability of DDMT, and also provide evidence to show that DDMT may explore a different set of candidate inputs due to the specific failure-revealing abilities of MT.

The differences between the total time costs of these two approaches are marginal.
Nevertheless, Perses$_{DDMT}$ consumes less time in 7 cases, while it requires relatively more time in one case.
\jmyrv{These results indicate that although the test function employed by DDMT requires more resources as the one utilized by \emph{ddmin},
the optimization of the entire reduction process may decrease the number of test invocations, and in turn leading to 
comparable time consumption to \emph{ddmin} for most of subjects.}

To summarize, Perses$_{DDMT}$ and Perses show comparable performance on the evaluated compiler subjects. 
These results confirm the practical applicability of DDMT in scenarios facing the test oracle problem.
Although Perses$_{DDMT}$ can be supported by benchmark-specific oracles in these scenarios, there remain difficulties in obtaining reliable and effective reference implementations.
Consequently, DDMT is more broadly applicable and effective for programs where reliable test oracles or comparable implementations are unavailable.

\section{Discussion}
\label{sec:dis}

\emph{Effectiveness and Cost Trade-off.}
Although DDMT achieves improvements in the size of resulting input and the number of queries, it incurs additional time overhead compared with \emph{ddmin}. The overhead mainly arises from the difference between the procedure of function \emph{mrtest} and that of function \emph{test}.
While the latter requires only one execution of the target program, the former involves the generation of the follow-up input, an additional execution of the program with the follow-up input, and the checking of the outputs from both executions.
Experimental results show that the additional time cost is often accompanied by a noticeable improvement in input size and the number of queries, indicating that the overhead is largely justified by the gained reduction effectiveness. 
Moreover, DDMT remains fully automated, reducing manual effort and improving scalability compared with human-intensive processes. 
Therefore, the results suggest that DDMT achieves a practical trade-off between reduction effectiveness and time efficiency.

\emph{Sensitivity to MRs.}
One limitation of DDMT is that its effectiveness and efficiency strongly depend on the quality of the employed MRs. 
As demonstrated in our experiments, different MRs lead to substantially different reduction results and reduction costs, further confirming the critical role of MR in DDMT. 
This limitation is inherited from MT itself, where the fault-detection capability is closely related to the strength and diversity of the selected MRs \cite{Liu2013}. 
Nevertheless, this issue can be alleviated in practice because a large number of effective MRs have already been proposed by the research community for diverse application domains, including compilers \cite{Le2014, xiao2022}, scientific software \cite{peng2021}, machine learning systems \cite{dwarakanath2018}, web applications \cite{chaleshtari2023}, and database systems \cite{lin2025}. 
Furthermore, prior studies have also investigated systematic methodologies and automated techniques for MR identification and generation \cite{li2025}. 
Therefore, although DDMT relies on MRs, the growing body of MR-related research provides strong support for its practical applicability.
Moreover, this observation also suggests that the effectiveness and efficiency of DDMT can be further enhanced via more powerful MRs.

\jmyrv{\emph{Threats to Validity.}}
Threats to internal validity may arise from the implementation of DDMT. 
To mitigate implementation-related errors, we adopted the publicly released package of the baseline approaches.
Moreover, we
carefully validated the implementation of MRs and the function \emph{mrtest}.
Another threat concerns the selection of subject programs. Our evaluation was conducted on a finite set of programs and inputs, which may not be fully representative of the broad range of software systems encountered in practice.
To reduce this threat, we included programs from both the Siemens suite and the benchmark of prior DD studies, across oracle-available and oracle-deficient scenarios.
Nevertheless, further experiments on larger and more diverse benchmark suites are needed to strengthen the generalizability of the findings.
Last but not least, the selected MRs may introduce selection bias. Since only a subset of all possible MRs can be evaluated, the observed superiority of certain MRs may not generalize to other unexamined MRs. 
To mitigate this threat, we collected MRs from existing literature and conducted comparisons analysis with respect to both the best and the worst MRs.

\section{Related Work}
\label{sec:rw}

\subsection{Delta Debugging Approaches}

Delta debugging \cite{Zeller2002} is a technique proposed by Zeller and Hildebrandt for automatically simplifying program inputs. 
The core algorithm of delta debugging is \emph{ddmin}, which can generate the smallest input that preserves the failure-revealing ability.
Subsequently, a series of approaches are proposed by enhancing and extending \emph{ddmin}. 
The Hierarchical delta debugging (HDD) \cite{Misherghi2016} exploits the input structure to minimize failure-causing inputs. 
Modernizing HDD \cite{hodovan2016} improves HDDs by proposing extended standard context-free grammars. 
Coarse HDD \cite{Hod2017} enables faster simplification of program inputs by filtering non-empty replacement fragments when visiting tree nodes, and it can also be used to accelerate simplification of original HDDs. 
HHDr \cite{kiss2018hddr} aims at improving the hierarchical minimization algorithm, which significantly reduces the time cost of getting the minimal result.
Perses \cite{Sun2018} exploits the formal syntax of programs and uses it to guide the reduction process, enabling effective and efficient reduction for general programs.

Motivated by the observation that traditional delta debugging explores the search space without exploiting information learned during the reduction process, ProbDD \cite{wang2021} adopts a statistical approach by maintaining probability distributions over components to predict which elements are more likely to be irrelevant.
It introduces a probabilistic model to estimate the likelihood that individual program elements belong to the failure-inducing set and continuously updates these estimates based on testing outcomes. 
Experimental results show that ProbDD can significantly reduce the number of required tests while preserving reduction effectiveness.
Driven by the strict grammatical constraints of target languages, Large language model-aided program reduction (LPR) \cite{zhang2024lpr} leverages LLMs to perform semantics-preserving transformations that expose additional reduction opportunities.
It combines iterative LLM-guided transformations with language-generic reduction and introduces a multi-level prompting mechanism to improve transformation accuracy. Evaluations show that LPR achieves substantially smaller reduced programs than existing state-of-the-art approaches while maintaining competitive efficiency.
WDD \cite{zhou2025wdd} addresses the limitation of prior delta debugging algorithms that treat program elements uniformly despite their varying sizes, resulting in suboptimal partitioning and reduction efficiency.
It introduces a weight-based partitioning strategy that assigns weights according to element size and guides the reduction process accordingly. Experimental results show that WDD improves minimization efficiency and can substantially reduce the number of required test executions compared with traditional delta debugging approaches.
In order to address the validity and efficiency limitations of traditional delta debugging approaches, GReduce \cite{ren2025} leverages the execution traces of test input generators to guide input reduction. It reduces generator executions instead of generated inputs directly, which ultimately preserves input validity and enables more effective and efficient reductions.

Instead of primarily targeting at enhancing the effectiveness and efficiency, this study focuses on extending the applicability of delta debugging to programs without test oracles. We demonstrated that the proposed approach, DDMT, is not only feasible, but also effective in producing simplified program inputs. 
DDMT is mainly designed for \emph{ddmin}, the core algorithm of delta debugging approaches. However, it should be obvious that DDMT can be similarly adapted to all of the delta debugging approaches discussed above.

\subsection{Integration of Metamorphic Testing With Other Methods}
MT has been integrated with various approaches, successfully extending the application of these approaches to programs without test oracles.
In light of the need of test oracles in fault-based testing, Chen et al. \cite{Chen01} proposed to integrate fault-based testing with MT, in order to enhance fault-based testing by alleviating the test oracle problem. 
To facilitate automated program debugging, Chen et al. \cite{Chen2011} proposed an MT-based approach, semi-proving, which is an integrated approach for program proving, testing and debugging.
Xie et al. \cite{Xie2013} proposed the concept of metamorphic slice, which combines program slicing with MT.
They further incorporated the metamorphic slice into spectrum-based fault localization, based on which the applicability of spectrum based fault localization has been significantly extended.
In the filed of automated program repair, Jiang et al. \cite{Jiang2016} proposed an integration of test suite based automated program repair with MT, which enables the application of automated program repair techniques without using test oracles, extending the application of automated program repair techniques to a broader domain.
Later, they further proposed an MT-based test suite construction method, for constructing effective input test suites for automated program repair \cite{Jiang2021}. 
In order to alleviate the oracle problem of combinatorial testing, Niu et al. \cite{niu2021} integrated MT into combinatorial testing such that the outcome of test cases can be automated checked.
The enhanced approach is able to be applied to more widely test scenarios.
To address the limited generalization of genetic programming in program synthesis under scarce labeled data, Sobania et al. \cite{sobania2023} integrated MT with genetic programming to validate program behavior via user-defined relations over unlabeled inputs.
The experimental results confirm the higher generalization rate of the integrated approach.

This study demonstrates a new integration paradigm for MT, and also provides a feasible solution for enhancing the applicability of delta debugging approaches.
On the other hand, apart from confirming that the integrated approach alleviates the test oracle problem, this study further reveals that the integration is of great benefit to the improvements in effectiveness and efficiency.

\section{Conclusion}
\label{sec:con}

In this paper, we proposed a novel delta debugging approach, DDMT, which seamlessly incorporates the technique of metamorphic testing (MT) into the reduction process of the \emph{ddmin} algorithm. 
With a newly designed MT-based test function, DDMT can be applied regardless of the availability of the test oracle, facilitating the application of delta debugging to a broader range of programs.
Experimental evaluation on the Siemens suite and C compilers confirms the feasibility of DDMT, demonstrating its applicability across both oracle-available and oracle-deficient scenarios.
Experimental results further show that although achieving better applicability, DDMT remains comparable to \emph{ddmin} in terms of both effectiveness and efficiency,  across a substantial number of subject programs and inputs.
In future work, we plan to further improve the effectiveness and efficiency of DDMT by incorporating more effective MRs.
We also intend to investigate the applicability of DDMT to large-scale software systems and more diverse testing scenarios.

\section*{DATA AVAILABILITY STATEMENT}
The implementation and benchmark of our work are publicly
available for experiment replication \cite{ddmt}. 

\ifCLASSOPTIONcaptionsoff
  \newpage
\fi

\bibliographystyle{IEEEtran}
\bibliography{IEEEabrv,Ref}

\end{document}